\documentclass[prb,twocolumn,showpacs,floatfix]{revtex4}
\usepackage{graphicx}
\usepackage{dcolumn}
\usepackage{bm}
\begin{document}
\title{Direct evidence for the magnetic ordering of Nd ions in NdMn$_2$Si$_2$ and  NdMn$_2$Ge$_2$ by high resolution inelastic neutron scattering}
\author{T. Chatterji$^1$, J. Combet$^1$, B. Frick$^1$, and A. Szyula$^2$ }
\address{$^1$Institut Laue-Langevin, 6 rue Joules Horowitz, BP 156, 38042 Grenoble Cedex 9, France\\
$^2$Institute of Physics, Jagellonian University, 30-059 Krakow, Reymonta 5, Poland
}
\date{\today}

\begin{abstract}
We have investigated the low energy nuclear spin excitations in NdMn$_2$Si$_2$ and  NdMn$_2$Ge$_2$ by high resolution inelastic neutron scattering. Previous neutron diffraction investigations gave ambiguous results about Nd magnetic ordering at low temperatures. The present element-specific technique gave direct evidence for the magnetic ordering of Nd ions. We found considerable difference in the process of the Nd magnetic ordering at low temperature in NdMn$_2$Si$_2$ and  NdMn$_2$Ge$_2$. Our results are consistent with those of magnetization and recent neutron diffraction measurements. 
\end{abstract}
\pacs{}
\maketitle
  The ternary compounds RT$_2$X$_2$ (R = rare earth; T = 3d, 4d or 5d transition metal and X= Si, Ge) exhibit interesting physical properties viz., heavy fermion behavior, superconductivity and exotic magnetism \cite{szytula89}. Most of these compounds crystallize in the body centered crystal structure of the ThCr$_2$Si$_2$ type (space group $I4/mmm$, no. 139). Although in most of these compounds the transition metal atoms do not carry any moment, the RMn$_2$X$_2$ compounds, however, carry magnetic moments on both the rare earth and the Mn atoms. Thus the presence of two magnetic sublattices and also the Mn-Mn, R-Mn and R-R competing exchange interactions in RMn$_2$X$_2$ lead to complex magnetic behavior at low temperatures. Here we concentrate our high resolution inelastic neutron scattering investigation on two of these Nd-based compounds NdMn$_2$Si$_2$ and NdMn$_2$Ge$_2$. During the last decades the magnetic properties of these compounds have been studied by magnetometric, NMR and neutron diffraction techniques \cite{narasimhan75,siek81,welter93,welter95,shigeoka88,tomka98} that lead to some contradictory conclusions. Narasimhan et al. \cite{narasimhan75} concluded that the Mn atoms in NdMn$_2$Si$_2$ and PrMn$_2$Si$_2$ order antiferromagnetically below $T_N = 380$ and $368$ K, respectively and also below 36 K, Nd sublattice becomes ordered and ferromagnetically coupled with the Mn sublattice. Siek et al. \cite {siek81} confirmed from their neutron diffraction measurements the higher temperature antiferromagnetic structure of these compounds, but reported that the R atoms do not order down to 1.8 K. Later neutron diffraction investigations \cite{welter93,welter95}, however, suggest that Nd atoms do order at low temperatures. It is to be noted that the neutron diffraction investigations often give ambigous results for magnetic structures containing two magnetic sub-lattices. In such situations additional element sensitive techniques can be very useful. Resonant X-ray magnetic scattering is one of such complementary technique and has already been successfully used in such situations. Here we use another less-well-known element-specific technique that can yield some useful information. This technique  \cite{heidemann70,heidemann72} uses high resolution inelastic neutron scattering that can measure hyperfine splitting of the nuclear levels of isotopes containing nuclear spins. We tested this  technique recently in several Nd-based compounds \cite{chatterji00,chatterji02,chatterji04,chatterji04a,chatterji08,chatterji08a,chatterji09,chatterji11}.
With this technique we measure the hyperfine interaction of the Nd atoms and probe the magnetic ordering of the Nd atoms only. The Mn nucleus   has spin but the spin dependent neutron scattering cross section \cite{sears99} of $^{55}$Mn is only $0.40 \pm 0.02$ barn compared to that of $^{143}$Nd which is $56 \pm 3$ barn. Therefore high-resolution inelastic scattering spectra give selective information of the magnetic ordering of Nd moments only.    
\begin{figure}
\resizebox{0.5\textwidth}{!}{\includegraphics{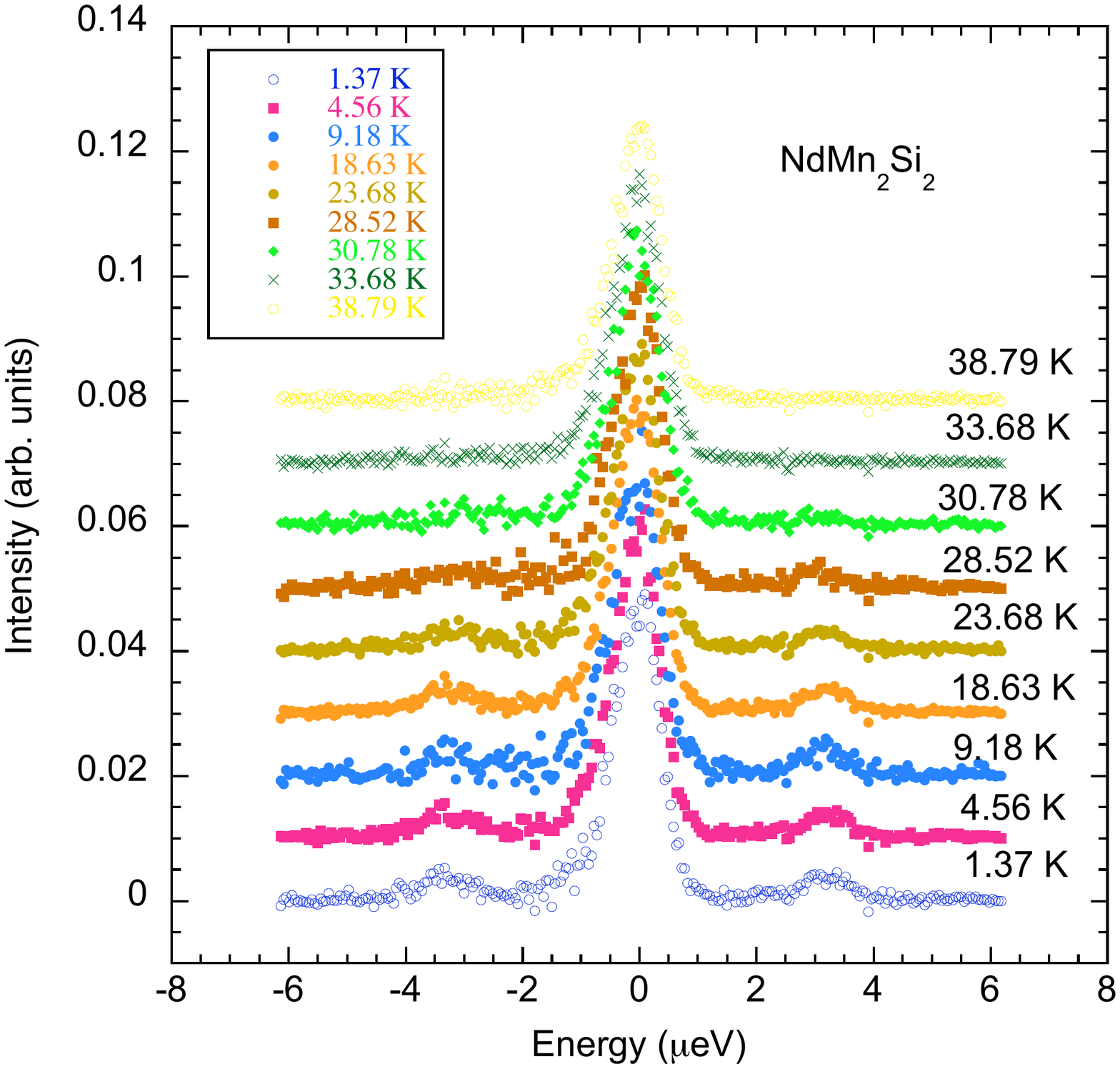}}

\caption { Energy spectra of NdMn$_2$Si$_2$ at several temperatures. The y-axes of the plots at T = 4.46, T = 9.18, T = 18.63 K, etc. have been shifted by 0.01, 0.02, 0.03 etc., respectively.
           } 
 \label{ndmnsispect1}
\end{figure}
\begin{figure}
\resizebox{0.5\textwidth}{!}{\includegraphics{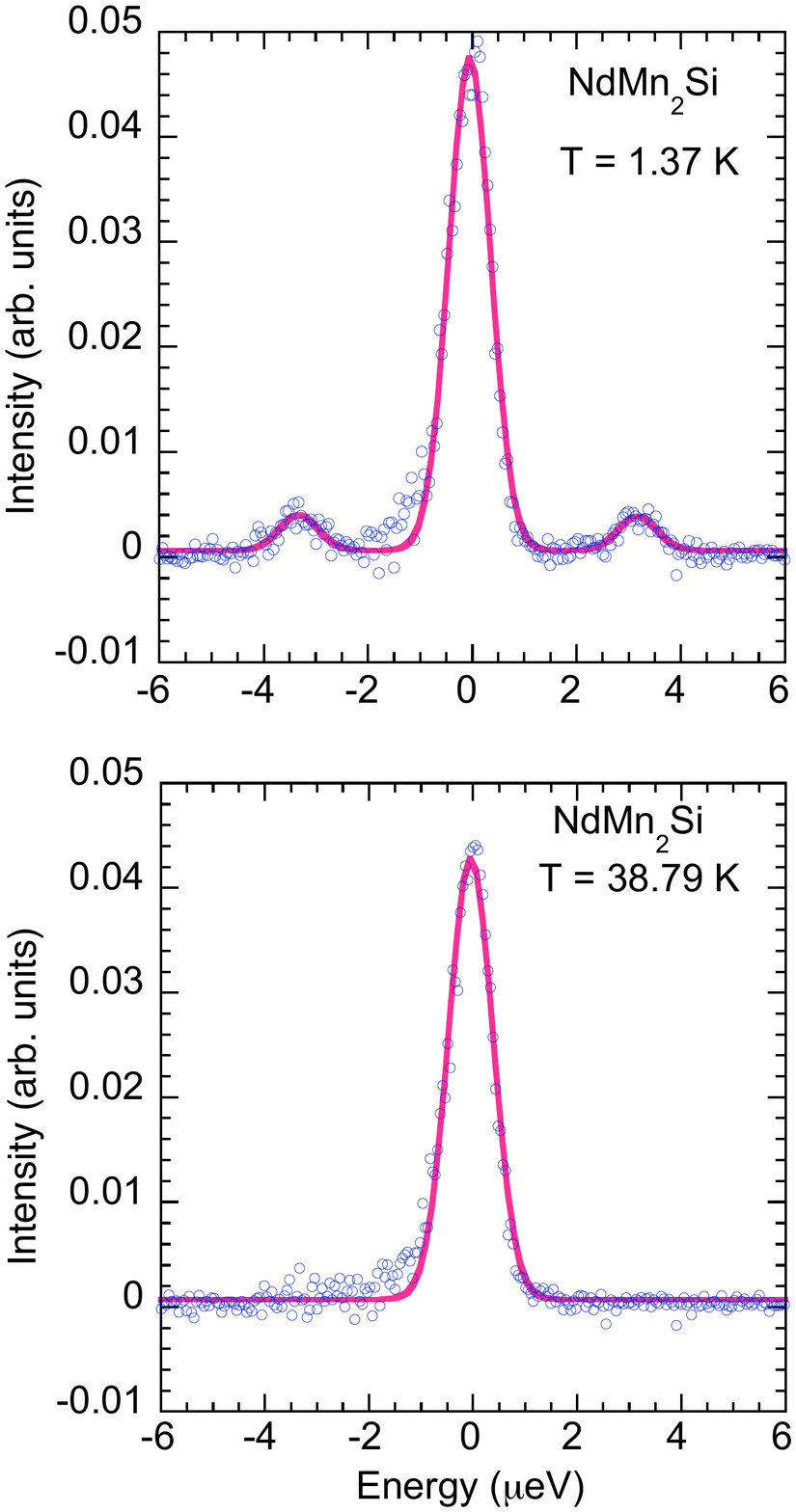}}

\caption { Energy spectra of NdMn$_2$Si$_2$ at $T = 1.37$ K and also at $T = 38.79$ K. The continuous curves are least squares fit with Gaussian functions.
           } 
 \label{ndmnsifit}
\end{figure}

\begin{figure}
\resizebox{0.5\textwidth}{!}{\includegraphics{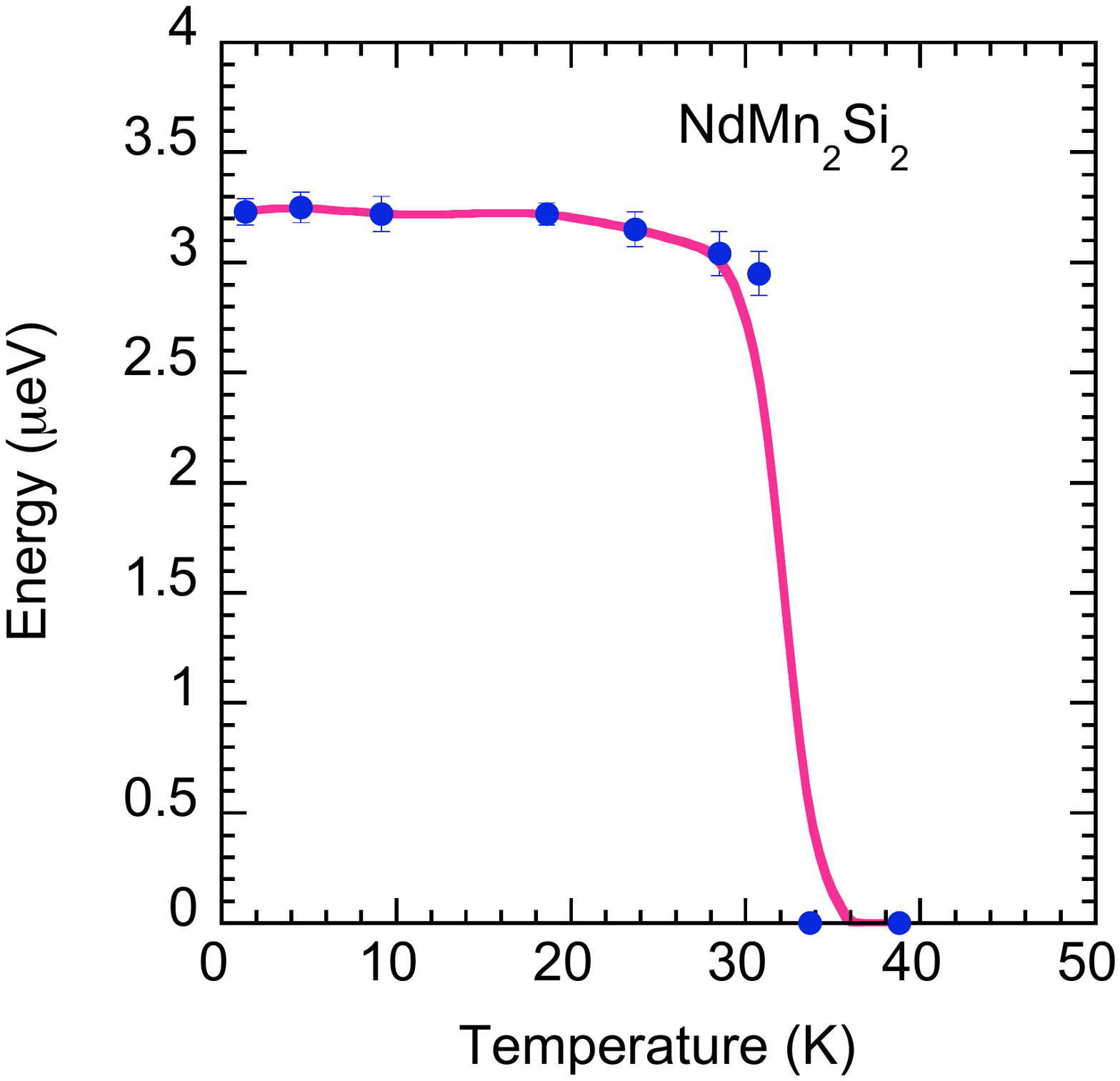}}
\caption {Temperature variation of the energy of the inelastic peak 
          of NdMn$_2$Si$_2$. The continuous curve is just a guide to the eyes. } 
\label{ndmnsiTdep}
\end{figure}

    The principle of the method of studying the hyperfine interaction by the inelastic neutron scattering
can be summarized \cite{heidemann70,heidemann72} as follows: If  neutrons with spin ${\bf s}$ 
are scattered from  nuclei with spins  ${\bf I}$, the probability that 
their spins will be flipped is $2/3$. The nucleus at which the 
neutron is scattered with a spin-flip, changes its magnetic quantum 
number $M$ to $M\pm 1$ due to the conservation of the 
angular momentum. If the nuclear ground state is split up into 
different energy levels $E_{M}$ due to the 
hyperfine magnetic field or an electric quadrupole interaction, then 
the neutron spin-flip produces a change of the ground state energy 
$\Delta E = E_{M} - E_{M\pm 1}$. This energy change is transferred 
to the scattered neutron. The double differential scattering cross section \cite{heidemann70} is 
given by  the following expressions:
\begin{equation}
	 \left(\frac{d^2\sigma}{d\Omega d\omega}\right)_
{inc}^{0}=\overline{(\overline{\alpha^{2}}-{\overline{\alpha}}^{2}+
\frac{1}{3}{\alpha^{\prime}}^{2}I(I+1))}e^{-2W(Q)}
\delta(\hbar\omega),
\label{heidemann01}
\end{equation}
\begin{equation}
\left(\frac{d^2\sigma}{d\Omega d\omega}\right)_
{inc}^{\pm}=
\frac{1}{3}\overline{{\alpha^{\prime}}^{2}I(I+1)}\sqrt{1\pm\frac{\Delta E}{E_{0}}}e^{-2W(Q)}
\delta(\hbar\omega\mp \Delta E)
\label{heidemann02}	 
\end{equation}
where $\alpha$ and $\alpha^{\prime}$ are coherent and spin-incoherent 
scattering lengths, $W(Q)$ is the Debye-Waller factor and $E_{0}$ is 
the incident neutron energy, $\delta$ is the Dirac delta function.  The long bar signs on top of equation (\ref{heidemann01}) and (\ref{heidemann02}) mean averages of the quantity on top of which they stay.
If the sample contains one type of isotope then 
$\overline{\alpha^{2}}-{\overline{\alpha}}^{2}$ is zero. Also 
$\sqrt{1\pm\frac{\Delta E}{E_{0}}} \approx 1$ because $\Delta E$ is 
usually much less than the incident neutron energy $E_{0}$. In this case 2/3 of 
incoherent scattering will be spin-flip scattering. The 
hyperfine splitting lies typically in the energy range of a few $\mu$eV. The inelastic spin-flip scattering of neutrons from the nuclear spins can 
yield this information provided the neutron spectrometer has the 
required resolution of about $1~\mu$eV or less and also the incoherent 
scattering of the nucleus is strong enough. Because of relatively large spin incoherent cross section of natural Nd, the Nd-based compounds are 
very much suitable for the studies of nuclear spin excitations. The Nd has the natural abundances of 12.18\% 
and 8.29\% of $^{143}$Nd and $^{145}$Nd isotopes, respectively. Both 
of these isotopes have nuclear spin of $I = 7/2$ and their incoherent 
scattering cross sections \cite{sears99} are relatively large, $55\pm 7$ and $5\pm 5$ barn
for $^{143}$Nd and $^{145}$Nd , respectively. Taking into account the natural abundance of these isotopes the total spin incoherent scattering of the natural Nd is $\sigma _i$(spin)$ = 7.1$ barn.
\begin{figure}
\resizebox{0.5\textwidth}{!}{\includegraphics{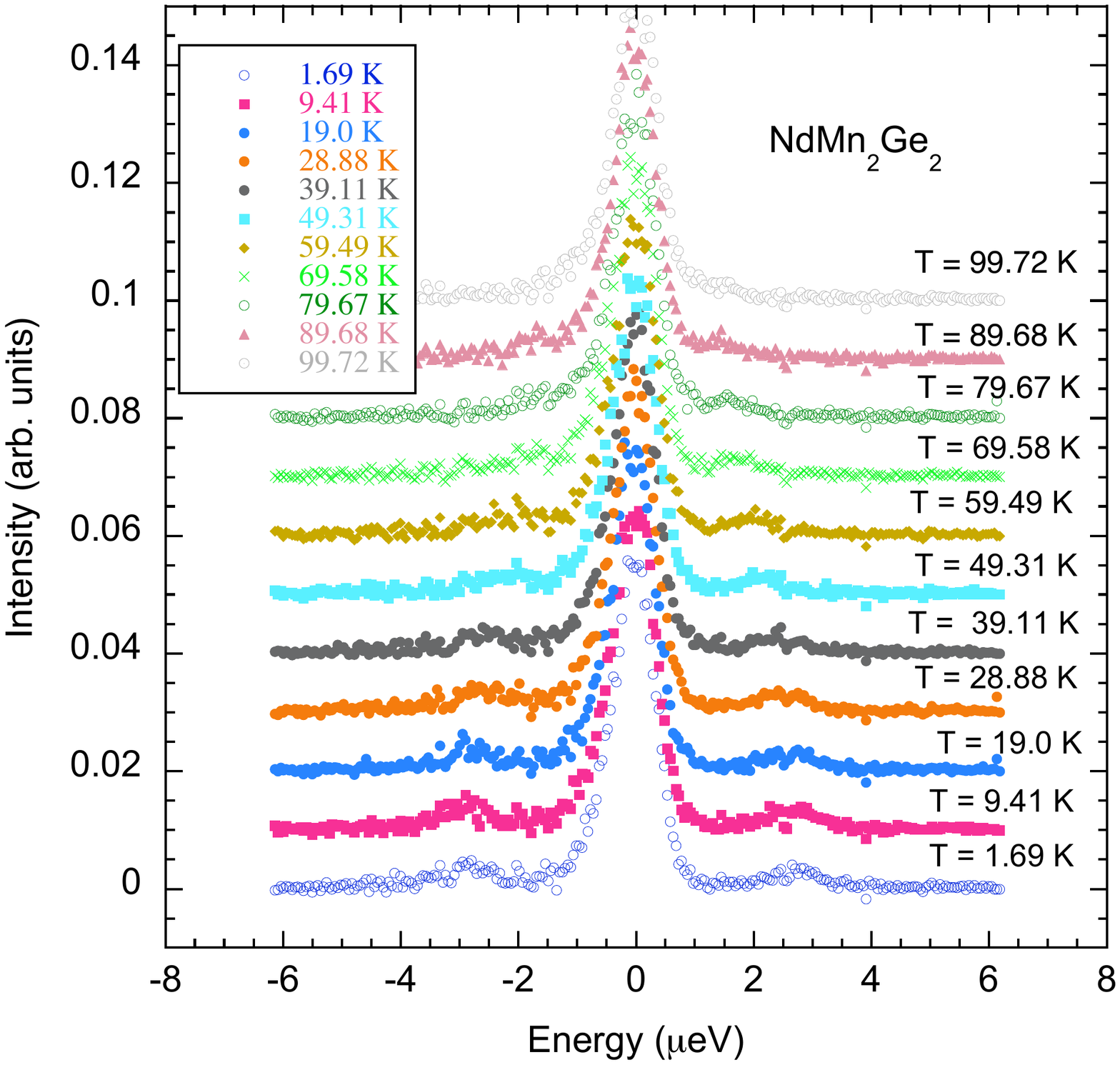}}

\caption { Energy spectra of NdMn$_2$Ge$_2$ at several temperatures. The y-axes of the plots at T = 9.41, T = 19.00, T = 28.88 K, etc. have been shifted by 0.01, 0.02, 0.03 etc., respectively.
           } 
 \label{ndmngespect1}
\end{figure}

\begin{figure}
\resizebox{0.5\textwidth}{!}{\includegraphics{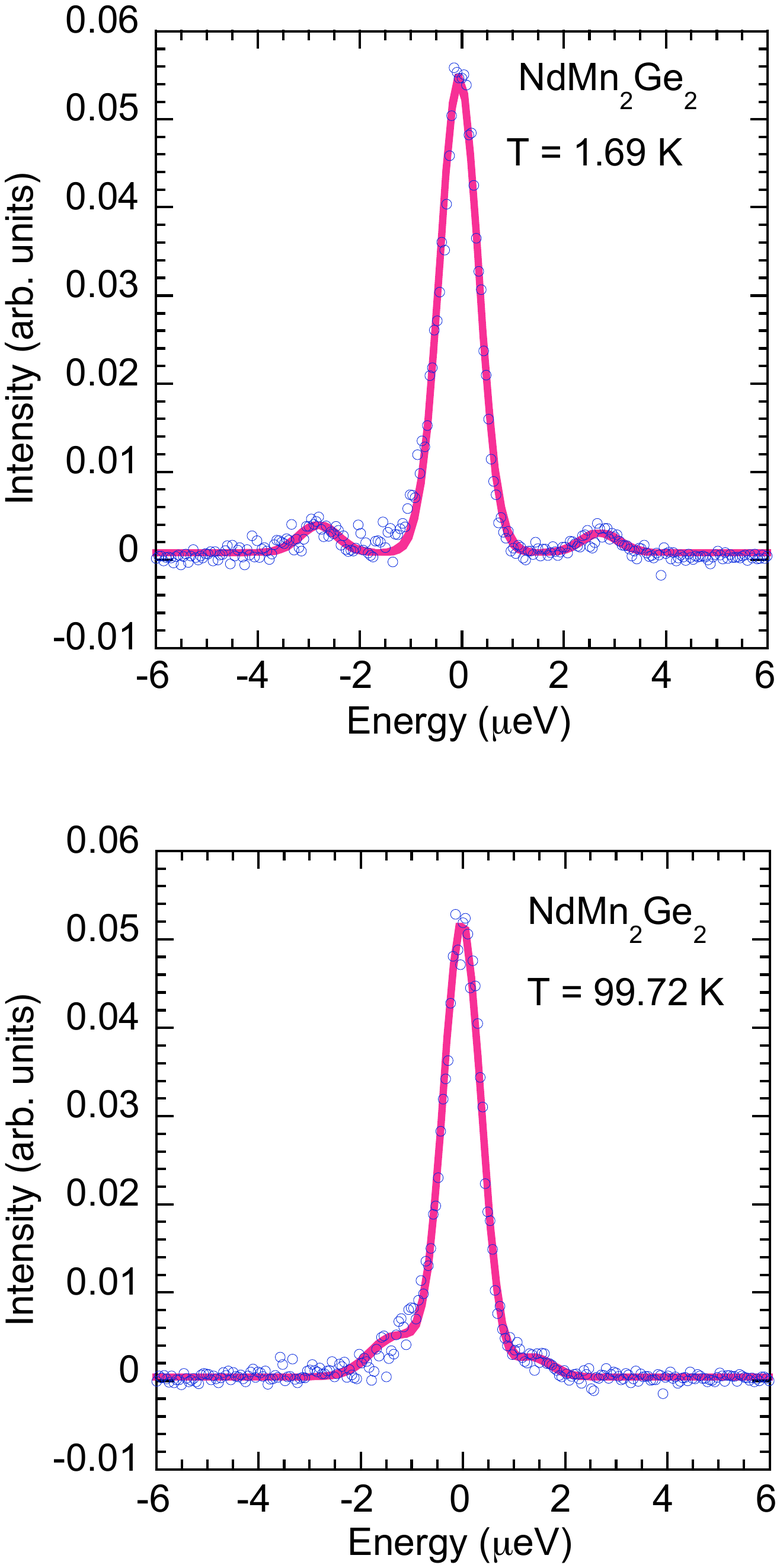}}

\caption { Energy spectra of NdMn$_2$Ge$_2$ at several temperatures. The continuous curves are least squares fit with Gaussian functions.
           } 
 \label{ndmngefit}
\end{figure}

\begin{figure}
\resizebox{0.5\textwidth}{!}{\includegraphics{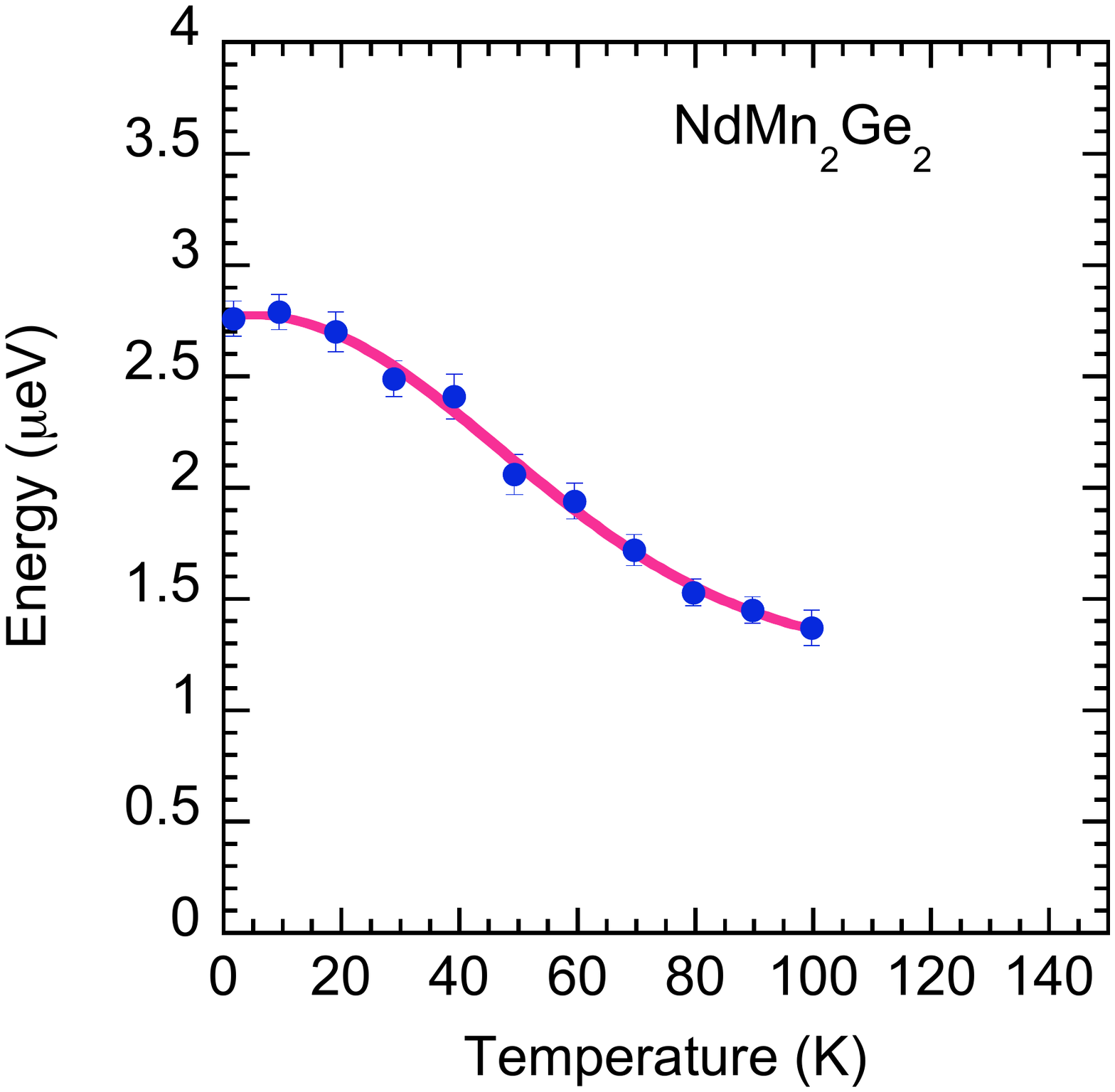}}
\caption {Temperature variation of the energy of the inelastic peak 
          of NdMn$_2$Ge$_2$. The continuous curve is just a guide to the eyes.  } 
\label{ndmngeTdep}
\end{figure}

\begin{figure}
\resizebox{0.5\textwidth}{!}{\includegraphics{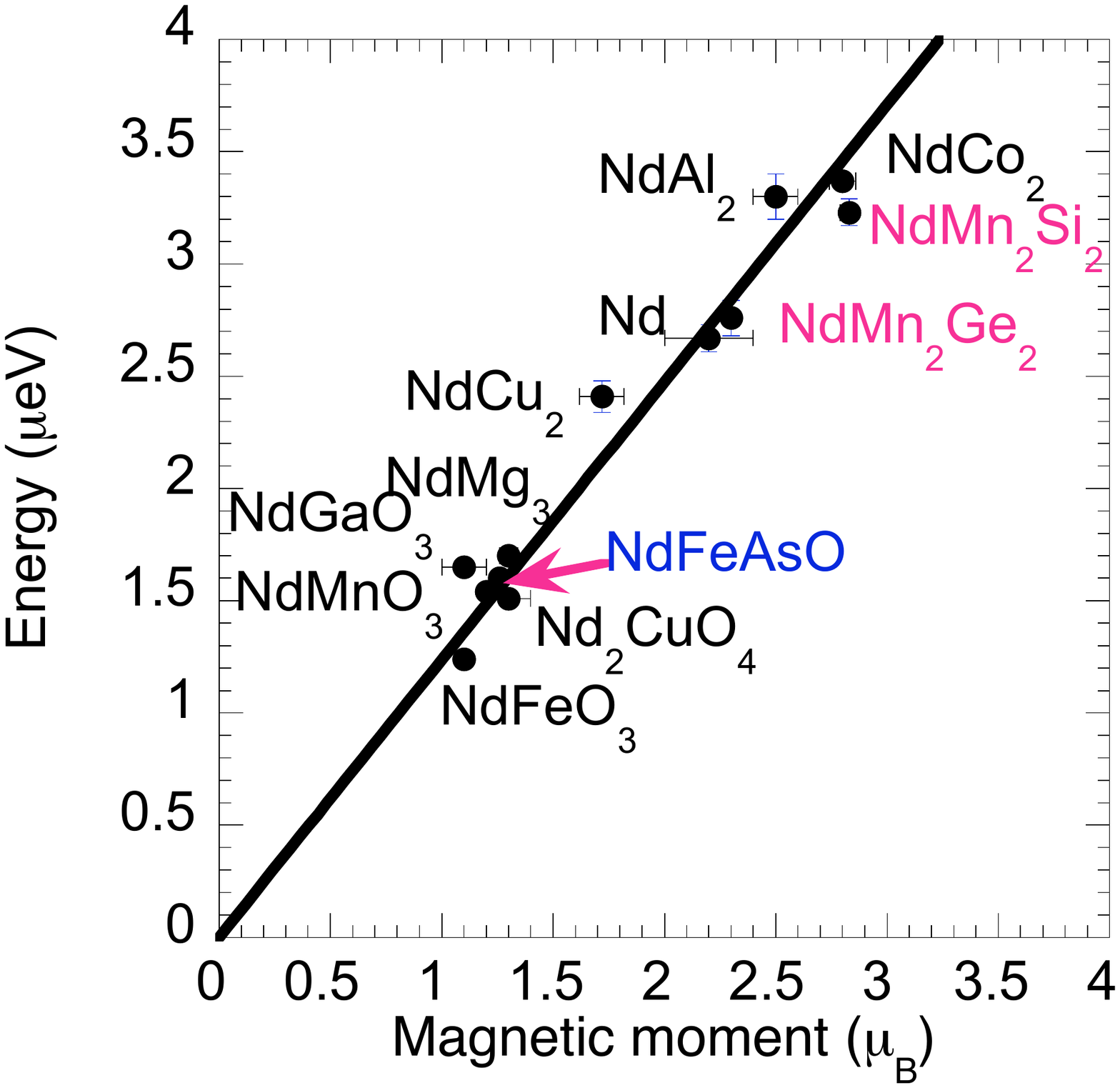}}
 \caption { Plot of energy of the inelastic signals in Nd$_2$CuO$_{4}$ \cite{chatterji00}, Nd 
metal \cite{chatterji02}, NdCu$_{2}$ \cite{chatterji04}, NdGaO$_{3}$ \cite{chatterji04a},  NdFeO$_3$ \cite{przenioslo06}, NdMg$_3$ \cite{chatterji08}, NdCo$_2$ \cite{chatterji08},
NdMnO$_3$ \cite{chatterji08a}  and NdAl$_2$ \cite{chatterji09}, NdFeAsO \cite{chatterji11}, NdMn$_2$Si$_2$ and NdMn$_2$Ge$_2$  vs. the corresponding 
electronic magnetic moment of Nd in these compounds determined by 
the refinement of the magnetic structure using magnetic neutron diffraction
intensities. The magnetic moment of Nd ion in NdFeAsO has been determined from the measured energy of the nuclear spin excitations \cite{chatterji11} by using the linear plot. } 
\label{ndmoment}
\end{figure}

NdMn$_2$X$_2$ (X=Si,Ge) samples were prepared by arc-melting under an argon atmosphere followed by annealing at $1073$ K for a one week in an evacuated quartz tube. The purity of the elements used for the samples preparation was 3N for Nd and Mn and 5N for Si and Ge. X-ray powder diffraction measurements with CuK$\alpha$ radiation indicate that both compounds are single phase and crystallize in the ThCr$_2$Si$_2$-type structure with the space group $I4/mmm$. About 5 g of NdMn$_2$Si$_2$ and NdMn$_2$Ge$_2$ powder samples were filled in an Al flat sample holder and were placed in the cryofurnace of the high resolution back scattering spectrometer IN10 of Institut Laue-Langevin in Grenoble.  Measurements were done on NdMn$_2$Si$_2$ and NdMn$_2$Ge$_2$ with the back-scattering spectrometer in the temperature range of about 2 - 40 and 2-100 K, respectively. 

Fig.~\ref{ndmnsispect1} shows the overview of the background-subtracted  inelastic spectra from NdMn$_2$Si$_2$  at several temperatures.  As expected there was no $Q$ dependence of the inelastic signals and therefore we plotted $Q$-integrated intensities  in Fig.~\ref{ndmnsispect1}. Inelastic signals are clearly visible at low temperatures and we interpret them to arise from the transition of the hyperfine split levels of nuclear spins of $^{143}$Nd and $^{145}$Nd isotopes. In order to determine the position, intensity and half-width of the peaks as a function of temperature we fitted the elastic and inelastic lines with Gaussian functions. Fig.~\ref{ndmnsifit} shows the least squares fits of the elastic and inelastic peaks of spectrum at $T = 1.37$ K with three Gaussian functions and at $T = 38.79$ K with a single Gaussian function. At $T = 1.37$ K clear inelastic signals were seen at $E = 3.23 \pm 0.03$ $\mu$eV on both energy loss and energy gain sides. The inelastic signals move only by small amounts to the central elastic peak at higher temperature.  However the signal becomes weaker at higher temperatures and could not be detected any longer above 28.5 K. We presume that the inelastic signals abruptly merge with the central elastic peak  at $T_{Nd} \approx 30$ K.  Fig. \ref{ndmnsiTdep} shows the temperature dependence of the energy of inelastic peaks. The low energy excitations in NdMn$_2$Si$_2$ is expected to be proportional the ordered magnetic moment of the Nd ions. The sharp increase of this energy below about $T_{Nd} \approx 34$ K indicates the ordering of the Nd magnetic moments. The temperature dependence of the low energy excitations in NdMn$_2$Si$_2$ is very similar to the temperature dependence of the neutron powder  diffraction intensities of the (001), (101) and (112) reflections observed by Welter et al. \cite{welter93}. The ordering temperature  $T_{Nd} \approx 34$ K of the Nd ions obtained  from the present investigation agrees well with that determined by Welter et al.  \cite{welter93} from the temperature dependence of the neutron diffraction intensity of the $(101)$ magnetic peak.

Fig.~\ref{ndmngespect1} shows the overview of the background-subtracted $Q$-integrated inelastic spectra from NdMn$_2$Ge$_2$  at several temperatures. Here also we see clearly the inelastic signals from the transition from the hyperfine-split nuclear levels. The data from NdMn$_2$Ge$_2$ have been treated in a similar way as those from NdMn$_2$Si$_2$ and Fig.~\ref{ndmngefit} shows the least squares fits of the elastic and inelastic peaks of spectrum at $T = 1.69$ K with three Gaussian functions and at $T = 99.72$ K with a single Gaussian function. Fig.~\ref{ndmngeTdep} shows the temperature variation of the energy of the inelastic signal. The energy varies continuously from the low temperature value of $E = 2.79 \pm 0.08$ $\mu$eV at $T = 1.69$ K to  to  $E = 1.37\pm 0.08$ $\mu$eV at $T = 99.72$ K, the highest temperature at which we measured the NdMn$_2$Ge$_2$ sample. Thus the ordered magnetic moment of the Nd ions in NdMn$_2$Ge$_2$ remains finite in the temperature range studied. The temperature variation is quite different from that obtained in NdMn$_2$Si$_2$ where a sharp increase of the energy was observed below about $T_{Nd} \approx 34$~K. However, the present results from NdMn$_2$Ge$_2$ is consistent with the results of NMR investigations of Tomka et al. \cite{tomka98} which show the presence of the ordered Nd magnetic moment up to about 150 K. Unfortunately due to limited neutron beam time our data are restricted up to about 100 K at which we could still see the weak inelastic signal albeit close to the central elastic peak. T0 get the quantitative data for the excitation energies above 100 K would be difficult due to the proximity of the inelastic signal to the central elastic peak.

We have studied hyperfine interactions  in several Nd-based compound recently by the high resolution neutron scattering technique and have found a linear relationship between the energy of nuclear spin excitations at low temperature and the ordered magnetic moment of the Nd ions in these compounds. One expects at low temperature in insulating Nd compounds an ordered magnetic moment of $3 \mu_B$ but due to the crystal-field effects one obtains a much lower ordered magnetic moments in neutron diffraction experiments. In Fig.~\ref{ndmoment} we show a plot of energy of inelastic peaks in Nd$_2$CuO$_{4}$ \cite{chatterji00}, Nd 
metal \cite{chatterji02}, NdCu$_{2}$ \cite{chatterji04}, NdGaO$_{3}$ \cite{chatterji04a},  NdFeO$_3$ \cite{przenioslo06}, NdMg$_3$ \cite{chatterji08}, NdCo$_2$ \cite{chatterji08}
NdMnO$_3$ \cite{chatterji08a}, NdAl$_2$ \cite{chatterji09},  NdFeAsO \cite{chatterji11}, NdMn$_2$Si$_2$ and NdMn$_2$Ge$_2$~vs.~the corresponding 
electronic magnetic moment of Nd ions in these compounds determined by 
the refinement of the magnetic structure using magnetic neutron diffraction
intensities. The data lie approximately on a straight line 
showing that the hyperfine field at the nucleus is approximately 
proportional to the electronic magnetic moment. The slope of the linear fit of the
data gives a value of $1.24 \pm 0.03 \mu eV$/$\mu_B$. It is to be noted that the data 
for the hyperfine splitting is rather accurate whereas the magnetic 
moments determined by neutron diffraction have large standard 
deviations and are dependent on the magnetic structure models. The magnetic 
structures are seldom determined unambiguously and the magnetic moment determined
from the refinement of a magnetic structure model is  relatively uncertain. In such cases the investigation of the low energy excitations
described here can be of additional help \cite{chatterji08a}. This is specially true for the complex magnetic structures
with two magnetic sub-lattices of which one sublattice contains Nd. Such complex magnetic  
structures, such as the parent compounds of newly discovered Fe-based superconductors, colossal magnetoresistive manganites and some multiferroic materials, are currently under intense
study. From the straight line plot (excluding NdMn$_2$Si$_2$ and NdMn$_2$Ge$_2$) we recently determined the ordered magnetic moment in  NdFeAsO \cite{chatterji11} for which no  unambiguous magnetic moment determination is done from neutron diffraction.  The magnetic moment determined from the linear plot is $1.26 \mu_B$. The determination of the magnetic moment of NdFeAsO was important because neutron diffraction investigation did not yield unambiguous value of magnetic moment.

In conclusion we have observed inelastic neutron scattering signals at low temperature in NdMn$_2$Si$_2$ and NdMn$_2$Ge$_2$, which give direct evidence for the magnetic ordering of Nd ions. The ordering of Nd magnetic moments at low temperature was controversial before.  Also we establish that the magnetic ordering of Nd sub-lattice in NdMn$_2$Si$_2$ and NdMn$_2$Ge$_2$ are quite different. In the former case the magnetic ordering of Nd is quite abrupt and seems spontaneous whereas in the latter case it is probably polarised by the ordering of Mn ion.

\end{document}